\documentclass[showpacs,preprintnumbers,amsmath,amssymb,superscriptaddress]{revtex4-1}
\usepackage[utf8]{inputenc}
\usepackage{amsmath,amssymb}
\usepackage{braket}
\usepackage{float}
\usepackage{bbm}
\usepackage{graphicx,graphics}
\usepackage{soul}
\usepackage{amsmath,amsfonts,amsthm,bm}
\usepackage{dsfont}
\usepackage{braket}
\usepackage{mathtools}
\usepackage{subcaption}
\usepackage{algorithm}
\usepackage{algorithmic}
\usepackage[normalem]{ulem}
\usepackage{amsbsy}
\usepackage{tikz}
\usepackage{cancel}
\usepackage{subcaption}
\usetikzlibrary{external}
\usetikzlibrary{automata,positioning}

\usepackage[normalem]{ulem} 
\usepackage{xcolor} 

\begin{document}

\title{Variational autoencoder reconstruction of complex many-body physics}
\author{I. Luchnikov}
\affiliation{Center for Energy Science and Technology, Skolkovo Institute of Science and Technology, 3 Nobel Street, Skolkovo, Moscow Region 121205, Russia}
\affiliation{Moscow Institute of Physics and Technology, Institutskii Per. 9, Dolgoprudny, Moscow Region 141700, Russia}
\author{A. Ryzhov}
\affiliation{Center for Energy Science and Technology, Skolkovo Institute of Science and Technology, 3 Nobel Street, Skolkovo, Moscow Region 121205, Russia}
\author{P.-J. C. Stas}
\affiliation{Department of Applied Physics, Stanford University 348 Via Pueblo Mall, Stanford, California 94305, USA}
\author{S. N. Filippov}
\affiliation{Moscow Institute of Physics and Technology, Institutskii Per. 9, Dolgoprudny, Moscow Region 141700, Russia}
\affiliation{Valiev Institute of Physics and Technology of Russian
Academy of Sciences, Nakhimovskii Pr. 34, Moscow 117218, Russia}
\affiliation{Steklov Mathematical Institute of Russian Academy of
Sciences, Gubkina St. 8, Moscow 119991, Russia}
\author{H. Ouerdane}\email{Corresponding author: h.ouerdane@skoltech.ru}
\affiliation{Center for Energy Science and Technology, Skolkovo Institute of Science and Technology, 3 Nobel Street, Skolkovo, Moscow Region 121205, Russia}

\begin{abstract}
Thermodynamics is a theory of principles that permits a basic description of the macroscopic properties of a rich variety of complex systems from traditional ones such as crystalline solids, gases, liquids, and thermal machines to more intricate systems such as living organisms and black holes to name a few. Physical quantities of interest, or equilibrium state variables, are linked together in equations of state to give information on the studied system, including phase transitions, as energy in the forms of work and heat, and/or matter are exchanged with its environment, thus generating entropy.  A more accurate description requires different frameworks, namely statistical mechanics and quantum physics to go deep into the microscopic properties of physical systems, and relate them to their macroscopic properties. These frameworks also allow to go beyond equilibrium situations. Given the notably increasing complexity of mathematical models to study realistic systems and their coupling to their environment that constrains their dynamics, both analytical approaches and numerical methods that build on these models, show limitations in scope or applicability. On the other hand, machine learning, i.e. data-driven, methods prove to be increasingly efficient for the study of complex quantum systems. Deep neural networks in particular have been successfully applied to many-body quantum dynamics simulations and to quantum matter phase characterization. In the present work, we show how to use a variational autoencoder (VAE) -- a state-of-the-art tool in the field of deep learning for the simulation of probability distributions of complex systems. More precisely, we transform a quantum mechanical problem of many-body state reconstruction into a statistical problem, suitable for VAE, by using informationally complete positive operator-valued measure. We show with the paradigmatic quantum Ising model in a transverse magnetic field, that the ground-state physics, such as, e.g., magnetization and other mean values of observables, of a whole class of quantum many-body systems can be reconstructed by using VAE learning of tomographic data, for different parameters of the Hamiltonian, and even if the system undergoes a quantum phase transition. We also discuss challenges related to our approach as entropy calculations pose particular difficulties.
\end{abstract}
\date{\today}


\maketitle

The empirical development of the dynamical theory of heat or classical equilibrium thermodynamics as we know it, was only possible because of the definition through a phenomenological approach of two fundamental physical concepts, which are the actual pillars of the theory: energy and entropy \cite{Muller2007}. It is with these two concepts that the laws (or principles) of thermodynamics could be stated and the absolute temperature be given a first proper definition. Though energy remains as fully enigmatic as entropy from the ontological viewpoint, the latter concept is not completely understood from the physical viewpoint. This of course did not preclude the success of equilibrium thermodynamics as evidenced not only by the development of thermal sciences and engineering, but also because of its cognate fields that owe it, at least partly or as an indirect consequence, their birth, from quantum physics to information theory. 

Early attempts to refine and give thermodynamics solid grounds started with the development of the kinetic theory of gases and of statistical physics, which in turn permitted studies of irreversible processes with the development of nonequilibrium thermodynamics \cite{Onsager1931,DeGroot1958,LeBellac2006,Apertet2014} and later on finite-time thermodynamics \cite{Andresen2011,Ouerdane2015} thus establishing closer ties between the concrete notion of irreversibility and the more abstract entropy, notably with Boltzmann's statistical definition \cite{Boltzmann1877} and Gibbs' ensemble theory \cite{Gibbs1902}. Notwithstanding conceptual difficulties inherent to the foundations of statistical physics such as, e.g., irreversibility and the ergodic hypothesis \cite{Penrose1979,Goldstein2019}, entropy acquired a meaningful statistical character and the scope of its definitions could be extended beyond thermodynamics, thus paving the way to information theory, as information content became a physical quantity per se, i.e. something that can be measured \cite{Shannon1948}. And, while quantum physics developed independently from thermodynamics, it extended the scope of statistical physics with the introduction of quantum statistics, led to the definition of the von Neumann entropy \cite{vonNeumann2018}, and also introduced new problems related to small, i.e. mesoscopic and nanoscopic, systems \cite{Datta1995,Heikilla2013}, down to nuclear matter \cite{Chomaz2004}, where the concepts of thermodynamic limit and ensuing standard definitions of thermodynamic quantities may be put at odds. 

Quantum physics problems that overlap with thermodynamics, are typically classified into different categories: ground state characterization \cite{bressanini2002robust}, thermal state characterization at finite temperature \cite{feiguin2005finite}, calculation of the dynamics of either closed or open systems \cite{carleo2012localization, chen2018dynamics}, state reconstruction from tomographic data \cite{lanyon2017efficient}, and quantum system control, which, given the complexity for its implementation, requires the development of new methods \cite{liao2019differentiable}. There are essentially two large families of techniques applicable to such problems: One is based on the quantum Monte Carlo (QMC) framework \cite{foulkes2001quantum}, which is powerful to overcome the curse of dimensionality by using the stochastic estimation of high-dimensional integrals; the other family encompasses methods that search solutions in the parametric set of functions, also called ansatz. The most used ansatzes are based on different tensor network architectures \cite{orus2014practical} as tensor-networks based methods show state-of-the-art performance for the characterization of one-dimensional strongly correlated quantum systems. One can solve either the ground-state problem by using the variational matrix product state (MPS) ground state search \cite{schollwock2011density}, or a dynamical problem using a time-evolving block decimation (TEBD) algorithm \cite{vidal2003efficient}. Quantum criticality of one-dimensional systems also can be studied by using a more advanced architecture called multi-scale entanglement renormalization ansatz (MERA) \cite{evenbly2013quantum}. The application of tensor networks is not restricted to one-dimensional systems, and one can describe an open quantum dynamics \cite{pollock2018non}, characterize the numerical complexity of an open quantum dynamics \cite{luchnikov2019simulation,Taranto2019}, perform tomography of non-Markovian quantum processes by using tensor networks \cite{Milz2018,luchnikov2019machine}, analyze properties of two dimensional quantum lattices by using projected entangled pair states (PEPS) \cite{verstraete2008matrix}, or solve classical statistical physics problems  \cite{levin2007tensor, evenbly2015tensor}. 

The cross-fertilization of quantum physics and thermodynamics has benefited much from the powerful quantum formalism and computational techniques; however, as thermodynamic concepts evolved from intuitive/phenomenological definitions, to classical-mechanics constructs extended with quantum physics and formalism when needed, thermodynamics, in spite of its undeniable theoretical and practical successes, never managed to fully mature into a genuine fundamental theory that firmly rests on strong basic postulates. This led a growing number of physicists to consider thermodynamics as incomplete on the one hand, and to think quantum theory as the underlying framework from which equilibrium and nonequilibrium thermodynamics emerge. Quantum thermodynamics \cite{Gemmer2009QThermo} is a fairly recent field of play where new ideas are tested while revisiting old problems related to cycles, engines and refrigerators, entropy production to name a few \cite{Allahverdyan2008,Thomas2011}. Further, quantum technology is a burgeoning field at the interface of physics and engineering, which seeks to develop devices able to harness quantum effects for computing and secure communication purposes. The wide scale development of such a kind of systems, which irreversibly interact with an infinite environment, rests on the ability to properly simulate the open quantum dynamics of their many-body properties and analyze coherence and dissipation at the quantum level. 

How fast quantum thermodynamics will progress is difficult to anticipate while unsolved problems are many, especially those related to the proper characterization of the physical processes, e.g., what qualifies as heat or work on ultra-short time and length scales where averages become irrelevant, is unclear and how the laws of thermodynamics may be systematically adapted still may be debated. To mitigate risks of slow progress, one may resort to approaches that do not rely on models of systems, but rather on data, the idea being to gain actual knowledge and understanding from data irrespective of how complex the studied system is. Machine learning (ML) provides perfectly suited tools for that purpose \cite{Bishop2006}. ML has a rather long history that can be dated back with the works of Bayes (1763) on prior knowledge that can be used to calculate the probability of an event as formulated by Laplace (1812). Much later (1913), Markov chains were proposed as a tool to describe sequences of events each being characterized by a probability of occurrence that depends on the actuality of the previous event only. And the main milestone is in 1950 with Turing's machine that can learn \cite{Turing1950}, shortly followed in 1951 by the first neural network machine \cite{Crevier1993}. Thanks to the huge increase in computational power over the last two decades, ML is now used for a wide variety of problems \cite{Bishop2006}, and quantum machine learning now shows extraordinary potential for faster and more efficient than ever treatment of complex quantum systems problems \cite{Biamonte2017}, one major challenge still residing in the development of the hardware capable to harness and transform this potentiality into actual tool. 

With the recent success in the field of deep learning, tools other than those based on tensor networks work as well as an ansatz. Restricted Boltzmann machine has been successfully applied as an ansatz to a ground state search, dynamics calculation and quantum tomography \cite{carleo2017solving,torlai2018neural,Tiunov2019} as well as convolution neural network to the two-dimensional frustrated $J_1-J_2$ model \cite{choo2019study}. The deep autoregressive model was applied very efficiently and elegantly to a ground state search of many-body quantum system and to classical statistical physics as well \cite{sharir2019deep, wu2019solving}. It was also recently shown how ML can establish and classify with high accuracy the chaotic or regular behavior of quantum billiards models and XXZ spin chains \cite{Kharkov2019}. Thus, it can be useful to transfer deep architectures from the field of deep learning to the area of many-body quantum systems. A variational autoencoder (VAE) was used for sampling from probability distributions of quantum states in \cite{Rocchetto2018}; in the present work we show that state-of-the-art generative architecture called conditional VAE can be applied to describe the whole family of the ground states of a quantum many-body system. For that purpose, using quantum tomography and reconstruction tools developed in \cite{carrasquilla2019reconstructing}, we consider the paradigmatic Ising model in a transverse-field as an illustration of the usefulness and efficiency of our approach. The use of VAE in such a problem is justified by the simplicity of VAE training, as well as its expressibility \cite{LeiWanglecture}.

The article is organized as follows. In Section 2, we give a brief recap of the physics of the Ising model in a transverse field. In Section 3, we develop our generative model in the framework of the tensor network. Section 4 is devoted to the variational autoencorder architecture. Results are shown and discussed in Section 5. The article ends with concluding remarks, followed a by a short series of appendices. 

\section{Transverse-field Ising model}
Among the rich variety of condensed matter systems, magnetic materials are a source of many fruitful problems whose studies and solutions inspired discussions and new models beyond their immediate scope. The Kondo effect (existence of a minimum of electrical resistivity at low temperature in metals due to the presence of magnetic impurities) is one such problem \cite{Hewson1993,Coleman2007} as it provides an excellent basis for studies of quantum criticality and absolute zero-temperature phase transitions \cite{Sachdev2000,Coleman2005} and also on a more fundamental level, a concrete example of asymptotic freedom \cite{Coleman2007}. Assuming infinite on-site repulsion, the single-impurity Anderson model \cite{Anderson1961,Hewson1993} permitted the establishment of a correspondence between Hamiltonian language and path integral for the development of non-perturbative methods in quantum field theory \cite{Fresard2007,Fresard2008}. One other important model is that of the Heisenberg Hamiltonian, defined for the study of ferromagnetic materials, and which, assuming a crystal subjected to an external magnetic field ${\bm B}$, reads \cite{Diu1996}: 

\begin{eqnarray}\label{Heisenberg}
H = -\sum_{\langle i,j\rangle}J_{ij} \hat{S}^{i} \hat{S}^{j} - {\bm h}\cdot\sum_j \hat{S}^j
\end{eqnarray}

\noindent where for ease of notations we introduced ${\bm h}=g\mu_B{\bm B}$, with $g$ being the Land\'e factor, and $\mu_B=e\hbar/2m_{\rm e}$ being the Bohr magneton ($e$: elementary electric charge, and $m_{\rm e}$: electron mass); $J_{ij}$ is a parameter that characterizes the nearest-neighbours exchange interaction between electron spins on the crystal sites $i$ and $j$ (the quantum spins $\hat{S}^{i}$ and $\hat{S}^{j}$ are vector operators whose components are proportional to the Pauli matrices). For simplicity, one may consider $J_{ij}\equiv J$ constant. If $J>0$ then the system is ferromagnetic and if $J<0$ the system is antiferromagnetic. Hereafter, we fix the electron's magnetic moment $g\mu_{\rm B}=1$.

Although Eq.~\eqref{Heisenberg} has a fairly simple form, the exact calculation of the partition function: 

\begin{equation}
    Z = {\rm Tr}~e^{-\beta H}
\end{equation}

\noindent where $\beta = 1/k_{\rm B}T$ is the inverse thermal energy, is possible on the analytical level with the mean-field approximation that simplifies the Hamiltonian \eqref{Heisenberg}, and also for one-dimensional systems, one difficulty of the Heisenberg Hamiltonian being that the 3 components of a spin vector operator do not commute. That said, Heisenberg's Hamiltonian is very useful to, e.g., study spin frustration \cite{Mila2015}, entanglement entropy \cite{Refael2004}, and also serve as a test case for density-matrix renormalization group algorithms \cite{Schollwock2005}. Under zero field, Heisenberg's Hamiltonian is also a simplified form of the Hubbard model at half-filling, thus including ferromagnetism in the scope of strongly correlated systems studies. 

A particular but very important approximation of Heisenberg's Hamiltonian and whose significance in physics, especially for the study of critical phenomena, cannot be underestimated is the so-called Ising model. In its initial formulation \cite{Ising1925}, Ising spins are $N$ classical variables, which may take $\pm 1$ as values, and form a one-dimensional (1D) system characterized by free or periodic boundary conditions. The classical partition function $Z$ may be calculated analytically for the 1D Ising model, and quantities such as the average total magnetization obtained directly \cite{Kramers1941}: 

\begin{equation}\label{magthermal}
   M = \frac{1}{\beta} \frac{\partial\ln Z}{\partial h}
\end{equation}

\noindent In the present work, we consider a 1D quantum spin chain whose Hilbert space is given by ${\cal H} = \bigotimes_i^N\mathbb{C}^2$. The system is described by the transverse-field Ising (TFI) Hamiltonian \cite{ovchinnikov2003antiferromagnetic}:

\begin{eqnarray}\label{IsingTF}
H = -J\sum_{\langle i,j\rangle} \sigma^i_z \sigma^{i+1}_z + h_x\sum_{i=1}^N \sigma^i_x.
\end{eqnarray}

\noindent where $\sigma^i_{\alpha}$ ($\alpha\equiv x,z$) is the Pauli matrix for the $\alpha$-component of the $i^{th}$ spin in the chain, and $h_x$ is the magnetic field applied in the transverse direction $x$. In this case, the spins are no longer the classical Ising ones and the two terms that compose the Hamiltonian $H$ do not commute, hence the need for a full quantum approach. An example of a real-world system that may be studied as a quantum Ising chain is cobalt niobate (CoNb$_2$O$_6$); in this case the spins that undergo the phase transition as the transverse field varies, are those of the Co$^{2+}$ ions \cite{Coldea2010}. The spin states are denoted $|+\rangle_i$ and $|-\rangle_i$ at ion site $i$. There are two possible ground states: when all $N$ spins are in the state $|+\rangle$, or in the state $|-\rangle$, i.e. when they are all aligned, which defines the ferromagnetic phase.

The phase transition from the ferromagnetic phase to the paramagnetic phase that we speak of now is of a quantum nature, and not of a thermal nature, as here it is driven only by the external magnetic field. More precisely, when the transverse field $h_x$ is applied with sufficient strength, the spins align along the $x$ direction, and the spin state at site $i$ is given as the superposition $\left(|+\rangle_i + |-\rangle_i\right)/\sqrt{2}$, which is nothing else but the eigenstate of the $x$-component of the spin. So, in this particular case, there is no need to raise the temperature of the system, initially in the ferromagnetic phase, beyond the Curie temperature, to make it a paramagnet: the many-body system remains in its ground state but its properties have changed. Further, it is interesting to note that unlike for the ferromagnetic phase, the quantum paramagnetic phase has spin-inversion symmetry. We recommend the reading of \cite{Sachdev2011} for an insightful discussion on quantum criticality. 

Now, we briefly comment on the quantity $\beta=1/k_{\rm B}T$ in the context of quantum phase transitions, which, strictly speaking, can only occur at temperature $T=0$ K. In fact, close to the absolute zero, where $\beta \rightarrow \infty$, their signatures can be observed as quantum fluctuations dominate thermal fluctuations in the criticality region, where the quantum critical point lies. The imaginary time formalism \cite{Matsubara1955}, where $\exp(-\beta H)$ is interpreted as an evolution operator, and the partition function $Z$ as a path integral, provides a way to map a quantum problem onto a classical one with the introduction of the imaginary time $\beta$ resulting from a Wick rotation in the complex plane, thus yielding one extra dimension to the model. In classical thermodynamics, to observe a phase transition in a system requires that its size (i.e. the number of constituents $N$) tends to infinity so that the order parameter is non-analytic at the transition point; so for the quantum transition, the thermodynamic limit entails the limit $\beta \rightarrow \infty$ also: the 1D TFI model is mapped onto an equivalent 2D classical Ising model \cite{Kogut1979}. The imaginary time formalism permits implementation of classical Monte Carlo simulations to study quantum systems. Further discussion, including the sign problem for the quantum spin-$1\!/\!2$ system, is available in \cite{LeBellac2006}. 

We have chosen the the transverse-field Ising model as an illustrative case for our study for several reasons. First, since this system is 1-dimensional, we can apply an MPS variational ground state solver \cite{schollwock2011density} and hence obtain the ground state solution in MPS representation. We can then perform fast and exact sampling for generation of large data sets for the training of the VAE. Next, this model can be solved analytically, which allows us to adequately benchmark our results. Finally, this model shows a nontrivial behavior around the quantum phase transition point at $h_x=1$, and thus constitutes an interesting example to apply a VAE.

\section{Generative model as a quantum state}
Many-body quantum physics is rich in high-dimensional problems. Often, however, with increasing dimensionality, these become extremely difficult or impossible to solve. One solving method is through the reformulation of the quantum mechanical problem as a statistical problem, when possible. This way, machine learning can be used to effectively solve such a problem, since machine learning is a tool for the solving of high-dimensional statistical problems \cite{krizhevsky2012imagenet}. Probabilistic interpretation allows for using powerful sampling-based methods that work efficiently with high dimensional data. 

An example of the reformulation of a quantum problem as a statistical problem is with informationally complete (IC) positive-operator valued measures (POVMs) \cite{holevo2011probabilistic}. POVMs describe the most general measurements of a quantum system. Each particular POVM is defined by a set of positive semidefinite operators $M^{\alpha}$, with the normalization condition $\sum_{\alpha}M^{\alpha}={\mathds 1}$, where ${\mathds 1}$ is the identity operator. The fact that the POVM is informationally complete means that using measurement outcomes one can reconstruct the state of a system with arbitrary accuracy. 

The probability of measurement outcome for a quantum system with the density operator $\rho$ is governed by Born's rule: $P[\alpha] = {\rm Tr}(\varrho M^{\alpha})$, where $\{M^{\alpha}\}$ is a particular POVM, and $\alpha$ is an outcome result. In other words, any density matrix can be mapped on a mass function, although not all mass functions can be mapped on a density matrix \cite{Filippov2010,appleby2017introducing}. Some mass functions lead to non-positive semidefinite ``density matrices'', which is not physically allowed. As such, quantum theory is a constrained version of probability theory. For a many-body system, these constraints can be very complicated, and direct consideration of quantum theory as a constrained probability theory is not fruitful. However, if one can access the samples of the IC POVM induced mass function, which is by definition physically allowed, this mass function can be reconstructed using generative modeling \cite{carrasquilla2019reconstructing, LeiWanglecture}. Samples can be obtained either by performing generalized measurements over the quantum system or by in silico simulation.

In the present work, we simulate measurements of the ground state of a spin chain with the TFI Hamiltonian, Eq.~(\ref{IsingTF}). As a local (one spin) IC POVM, we use the so-called symmetric IC POVM for qubits (tetrahedral) POVM \cite{Caves1999}:

\begin{eqnarray}
&&M^{\alpha}_{\rm tetra} = \frac{1}{4}\left({\mathds 1} + \bm{s^{\alpha}}\bm{\sigma}\right),\ \alpha \in (0, 1, 2, 3), \ \bm{\sigma} = \left(\sigma_x, \sigma_y, \sigma_z\right),\nonumber\\
&&s^0 = (0, 0, 1), \ s^1 = \left(\frac{2\sqrt{2}}{3}, 0, -\frac{1}{3}\right), \ s^2 = \left(-\frac{\sqrt{2}}{3}, \sqrt{\frac{2}{3}}, -\frac{1}{3}\right), \ s^3 = \left(-\frac{\sqrt{2}}{3}, -\sqrt{\frac{2}{3}}, -\frac{1}{3}\right).
\end{eqnarray}

\noindent Note that many-spins generalization of local IC POVM can easily be obtained by considering the tensor product of local ones:

\begin{eqnarray}
M^{\alpha_1, \dots, \alpha_N}_{\rm tetra} = M^{\alpha_1}_{\rm tetra} \otimes M^{\alpha_2}_{\rm tetra}\otimes \dots \otimes M^{\alpha_N}_{\rm tetra}.
\end{eqnarray}

In order to simulate measurements outcome under the IC POVM described above, we implement the following numerical scheme: First, we run a variational MPS ground state solver to obtain the ground state of the TFI model in the MPS form:

\begin{equation}
\Omega_{i_1, i_2, \dots, i_N}=\sum_{\beta_1, \beta_2, \dots, \beta_{N-1}}A^{1}_{i_1\beta_1}A^{2}_{\beta_1 i_2 \beta_2}\dots A^{N}_{\beta_{N-1}i_N} 
\end{equation}

\noindent where we use the tensor notation instead of the bra-ket notation for further simplicity, and we obtain the MPS representation of IC POVM induced mass function:

\begin{eqnarray}
&&P[\alpha_1, \alpha_2, \dots, \alpha_N] = \sum_{\delta_1, \delta_2,\dots,\delta_{N-1}} \pi_{\alpha_1\delta_1}\pi_{\delta_1\alpha_2\delta_2}\dots\pi_{\delta_{N-1}\alpha_N},\nonumber\\
&&\pi_{\delta_{n-1}\alpha_n\delta_n}=\pi_{\underbrace{\beta_{n-1}\beta_{n-1}'}_{{\rm multi-index}\ \delta_{n-1}}\alpha_n\underbrace{\beta_n\beta_n'}_{{\rm multi-index}\ \delta_{n}}}=\left[M_{\rm tetra}\right]^{\alpha_n}_{ij}A^n_{\beta_{n-1}j\beta_n}\left[A^{n}\right]^*_{\beta_{n-1}'i\beta_n'}
\end{eqnarray}

\begin{figure}[ht]
    \centering
    \includegraphics[scale=0.7]{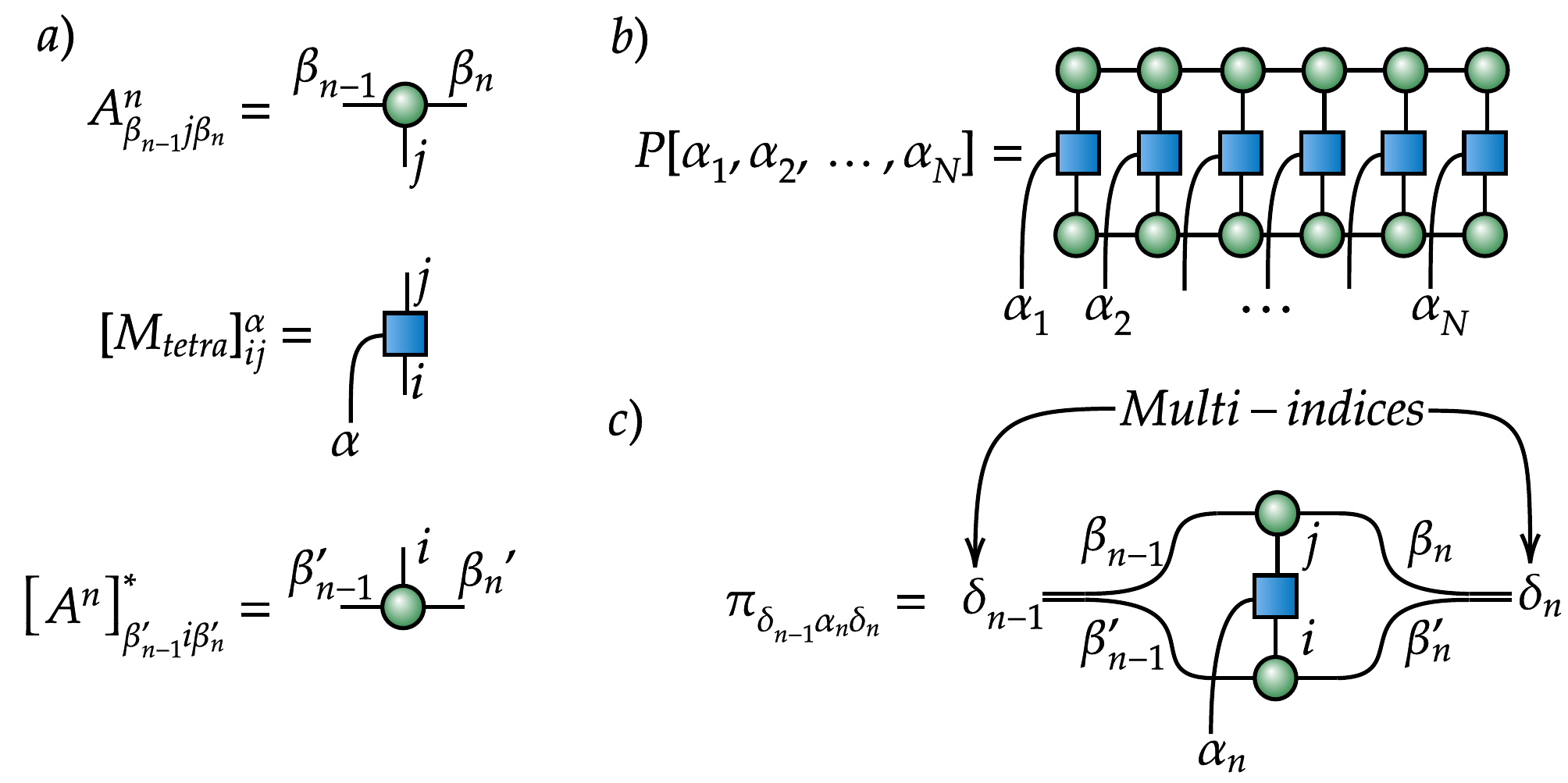}
    \caption{Tensor diagrams for a) building blocks b) MPS representation of measurement outcome probability and c) its sub-tensor.}
    \label{fig1}
\end{figure}
\begin{figure}[ht]
    \centering
    \includegraphics[scale=0.7]{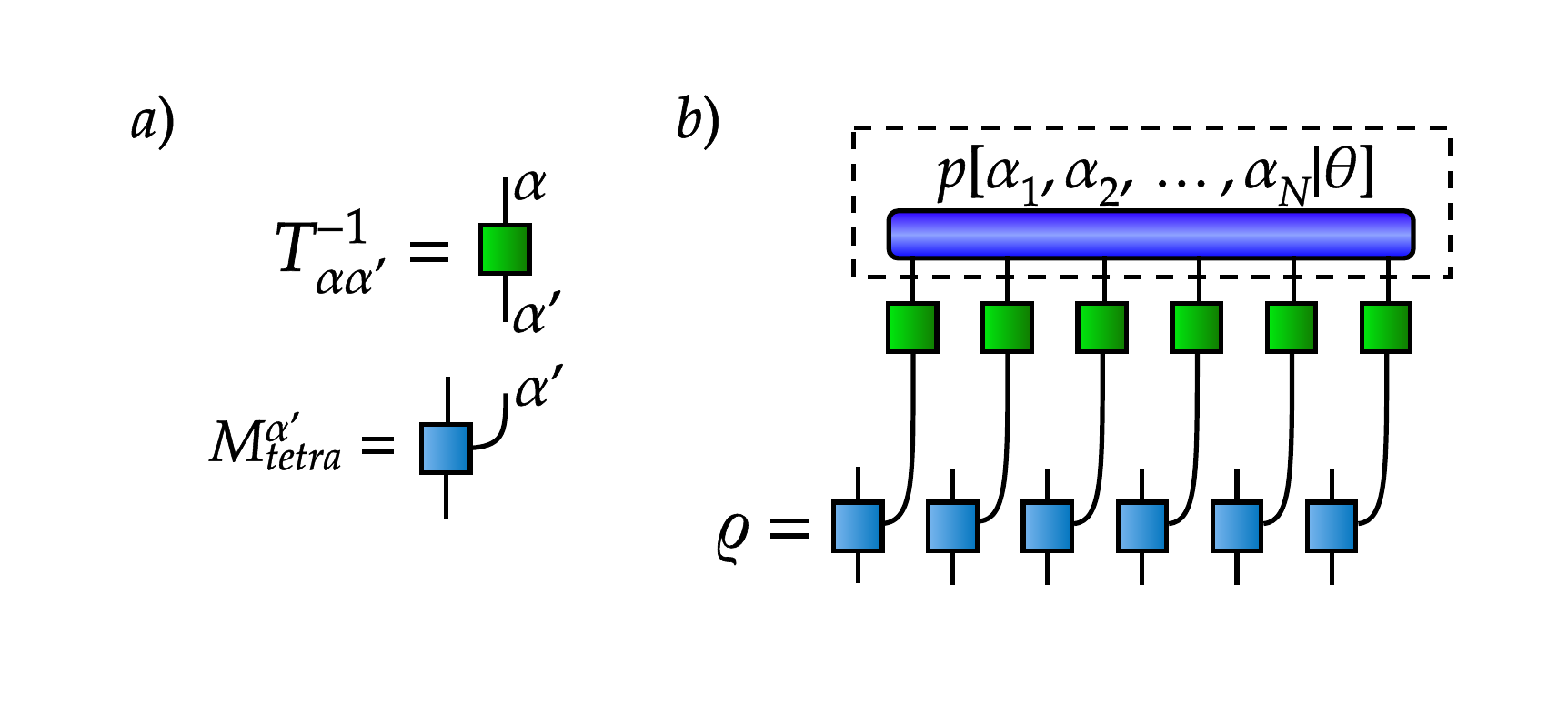}
    \caption{Tensor diagrams for a) building blocks b) inverse transformation from a mass function to a density matrix.}
    \label{fig2}
\end{figure}


\noindent whose diagrammatic representation \cite{orus2014practical} is shown in Fig.~\ref{fig1}. Next, we produce a set of samples of size $M$: $\{\alpha_1^i, \alpha_2^i,\dots,\alpha_N^i\}_{i=1}^M$ from the given probability. The sampling can be efficiently implemented as shown in Appendix B. We call this set of samples (outcome measurements) a data set, which may then be used to train a generative model $p[\alpha_1, \alpha_2,\dots,\alpha_N|\theta]$ to emulate the true mass function $P[\alpha_1, \alpha_2,\dots,\alpha_N]$. Here $\theta$ is the set of parameters of the generative model, which is trained by maximizing the logarithmic likelihood ${\cal L}(\theta)=\sum_{i=1}^M\log p[\alpha_1^i, \alpha_2^i,\dots,\alpha_N^i|\theta]$ with respect to the parameters $\theta$ \cite{myung2003tutorial}. The trained generative model fully characterizes a quantum state. The density matrix is obtained by applying an inverse transformation to the mass function \cite{Filippov2010b}:

\begin{eqnarray}
&&\varrho = \sum_{\alpha_1, \alpha_2, \dots, \alpha_N}p[\alpha_1,\alpha_2,\dots,\alpha_N|\theta][M^{\alpha_1}_{\rm tetra}]^{-1}\otimes [M^{\alpha_2}_{\rm tetra}]^{-1} \otimes\dots\otimes [M^{\alpha_N}_{\rm tetra}]^{-1},\nonumber\\ &&[M^{\alpha}_{\rm tetra}]^{-1} = \sum_{\alpha'}T^{-1}_{\alpha\alpha'}M^{\alpha'}_{\rm tetra},\nonumber\\ 
&&T_{\alpha\alpha'} = {\rm Tr}\left(M^{\alpha}_{\rm tetra}M^{\alpha'}_{\rm tetra}\right),
\end{eqnarray}

\noindent the diagrammatic representation of which is given in Fig.~\ref{fig2}. Note that the summation included in the density matrix representation is numerically intractable, but we can estimate it using samplings from the generative model.

\begin{figure}[ht]
    \centering
    \includegraphics[scale=0.7]{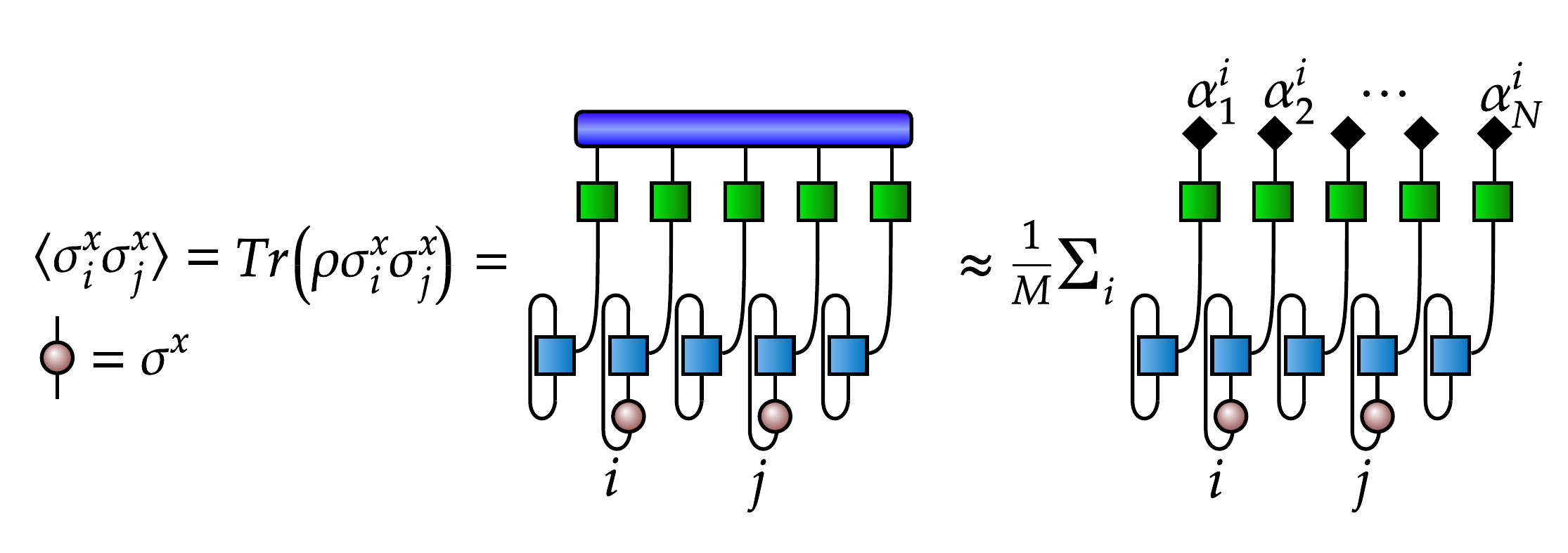}
    \caption{Tensor diagrams representing calculation of two-point correlation function.}
    \label{fig3}
\end{figure}

Our goal is to use a generative model as an effective representation of quantum states to calculate the mean values of observables such as , e.g., two-point and higher-order correlation functions. An explicit expression of the two-point correlation function obtained by sampling from the trained generative model is shown in Fig.~\ref{fig3}. To obtain the ground state of the TFI model we use a variational MPS ground state search and we pick the bond dimension of MPS equal to $25$ and perform $5$ DMRG sweeps to get an approximate ground state in the MPS form. We use the variational MPS solver provided by the mpnum toolbox \cite{mpnum}.

\section{Variational Autoencoder Architecture}

In our work we use a conditional VAE \cite{sohn2015learning} to represent quantum states. A conditional VAE is a generative model expressed by the following probability distribution: 

\begin{eqnarray}
p[x | \theta, h] = \int p[x | z, \theta, h] p[z] dz,
\end{eqnarray}

\noindent where $x$ is the data we want to simulate, $\theta$ represents the VAE parameters, which can be tuned to get the desired probability distribution over $x$, $h$ is the condition, and $z$ is a vector of latent variables. In our case $x$ is the quantum measurement outcome in one-hot notation. A collection of measurement outcomes is a matrix of size $N \times 4$, where $N$ is the number of particles in the chain, and 4 is the number of possible outcomes of the tetrahedral IC POVM, which is either $[1 0 0 0]$, $[0100]$, $[0010]$, or $[0001]$. $h$ is the external magnetic field. The probability distribution $p[x | z, \theta, h]$ can thus be written as:

\begin{eqnarray}
p[x | z, \theta, h] = \prod_{i=1}^N \prod_{j=1}^4 \pi_{ij} (z, h, \theta)^{x_{ij}},
\end{eqnarray}

\noindent where $\pi_{ij}(z, h, \theta)$ is the neural network in our architecture; and more precisely, $\pi_{ij}$ is the probability of the $j^{th}$ outcome of the POVM for the $i^{th}$ spin with $\sum_{j=1}^N \pi_{ij} = 1$ and $\pi_{ij} \geq 0$. The quantity $p[z]$ is the prior distribution over latent variables, which is simply given by ${\cal N}(0,I) =\frac{1}{\sqrt{2\pi}^N}\exp\left\{-\frac{1}{2}z^{\rm T}z\right\}$, with $I$ being the identical covariance matrix. We take the number of latent variables equal to the number of spins, $N$. Essentially, we want to optimize our VAE so that its probability matches the probability of the quantum measurement outcomes as closely as possible. This can be done using the well-known maximum likelihood estimation: 

\begin{eqnarray}
\theta_{MLE} = \underset{\theta}{\rm argmax}\sum_{i=1}^M \log(p[x_i | \theta, h]),
\label{eq:argmax}
\end{eqnarray}

\noindent where $\{x_i\}_{i=1}^M $ is the data set of outcome measurements. We cannot simply maximize this function using, for example, a gradient descent method, due to the presence of hidden variables in the structure of this function. However, we can overcome this problem by using the Evidence Lower Bound (ELBO) \cite{kingma2013auto} and the reparametrization trick shown in \cite{rezende2014stochastic}. The detailed description of the procedure is given in the Appendix A.

\begin{figure}
    \centering
    \includegraphics[scale=0.4]{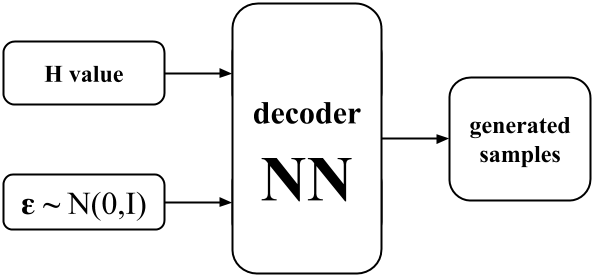}
    \caption{Sampling scheme with the trained VAE}
    \label{fig:sampling_architecture}
\end{figure}

Once trained, the VAE is a simple and efficient way to produce new samples from its probability distribution. It can be done in three steps. First, we produce a sample from the prior distribution $p[z] = {\cal N}(0, I)$. Next, we feed this sample and the external magnetic field value into the neural network decoder $\pi_{ij}(z, \theta, h)$, which returns the matrix of probabilities. Finally, we sample from the matrix of probability $\pi_{ij}(z, \theta, h)$ to generate ``fake'' outcome measurements. A visual representation of the sampling method is shown in Fig. \ref{fig:sampling_architecture}.

In many problems, gradients of observables with respect to different model parameters yield quantities of interest. For example, one may consider the magnetic differential susceptibility tensor $\chi_{i j}=\partial \mu_i/\partial h_j$. It can be done efficiently by using backpropagation through the VAE architecture but, as samples from the VAE are discrete, a straightforward backpropagation is impossible. In recent papers \cite{jang2016categorical, kusner2016gans, maddison2016concrete}, a method called the Gumbel-softmax, was introduced to overcome this difficulty through continuous relaxation. The spirit, and hence the physical meaning of the method, may be understood with a short discussion of the so-called simulated annealing technique, which is often used to solve discrete optimization problems. Broadly speaking, the simulated annealing rests on the introduction of a parameter that acts as an artificial ``temperature'', which varies continuously to modify the state of the system in search of a global optimum. Starting from a given state, for some values of the temperature, if the system mostly explores the neighboring states, moving among them and possibly in the vicinity of the ``better'' ones, i.e. with lower energy, it may get and remain close to a local optimum, or local energy minimum in the thermodynamic language; but to avoid remaining in a locally optimal region, ``bad'' moves leading to worse (i.e. higher energy) states are useful to explore the temperature space more completely improving the chance to find a global optimum or at least to be near it. To each move an energy variation, $\Delta E$, is associated. It is the continuous character of the fictitious temperature that makes the discrete problem continuous as the probability $\exp(-\Delta E)/k_{\rm B}T$ of acceptance of a state is continuous. Although this approach has been known for a long time \cite{Metropolis1953}, it remains topical and under active development \cite{RevModPhys.80.1061,yavorsky2019highly}. The method of continuous relaxation we use also exploits such an artificial temperature to make discrete samples continuous.

The Gumbel-softmax trick, consists of three steps:

\begin{enumerate}
    \item We calculate the matrix of log probabilities, taking element-wise logarithm of decoder network output:\\ $\log \Pi = \begin{bmatrix} \log \pi_{11} & \log \pi_{12} \dots \log \pi_{1N}\\ \log \pi_{21} & \log \pi_{22} \dots \log \pi_{2N} \\ \log \pi_{31} & \log \pi_{32} \dots \log \pi_{3N} \\ 
    \log \pi_{41} & \log \pi_{42} \dots \log \pi_{4N}\end{bmatrix}$,
    \item We generate a matrix of samples from the standard Gumbel distribution $G$ and sum it up element wise with the matrix of log probabilities $\log \Pi$: $Z = \log \Pi + G$,
    \item Finally we take the ${\rm softmax}$ function of the result from the previous step: $x_{\rm soft}^{\rm fake}(T) = {\rm softmax}(Z / T)$, where $T$ is a temperature of softmax. The softmax functions is defined by the expression: ${\rm softmax}(x_{ij})=\frac{\exp\left(x_{ij}\right)}{\sum_{i}\exp\left(x_{ij}\right)}$.
\end{enumerate}

\noindent The quantity $x_{\rm soft}^{\rm fake}(T)$ has a number of remarkable properties: first, it becomes an exact one-hot sample when $T \rightarrow 0$; second, we can backpropagate through soft samples for any T$>0$. The method is validated in the next section.

\section{Results}

\begin{figure}[h!]
\centering
\includegraphics[scale=0.4]{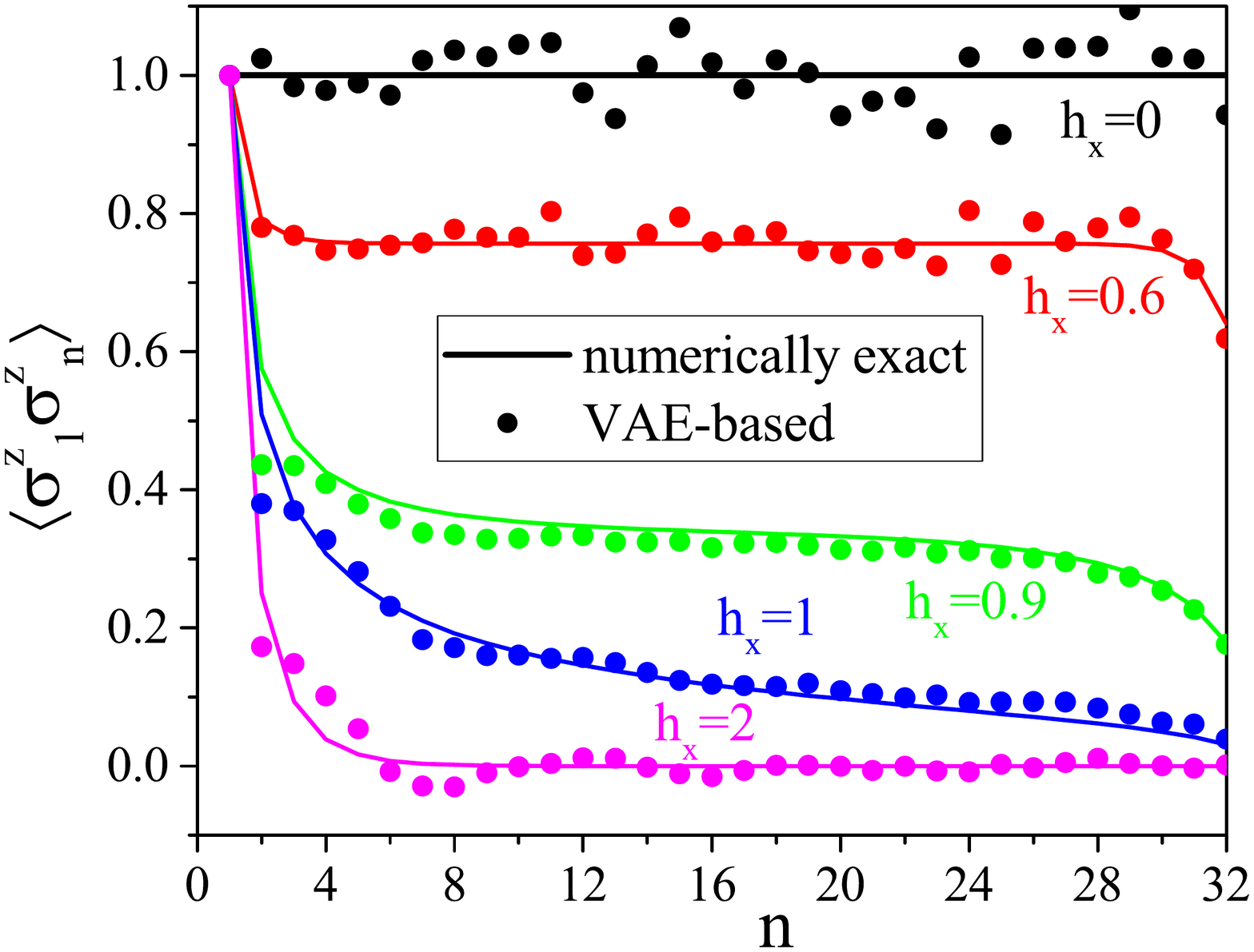}
\caption{ Two-point correlation function $\langle \sigma^z_1 \sigma^z_n \rangle$ for different values of external magnetic field $h_x$. }
\label{fig:corr_zz}
\end{figure}

\begin{figure}[h!]
\centering
\includegraphics[scale=0.4]{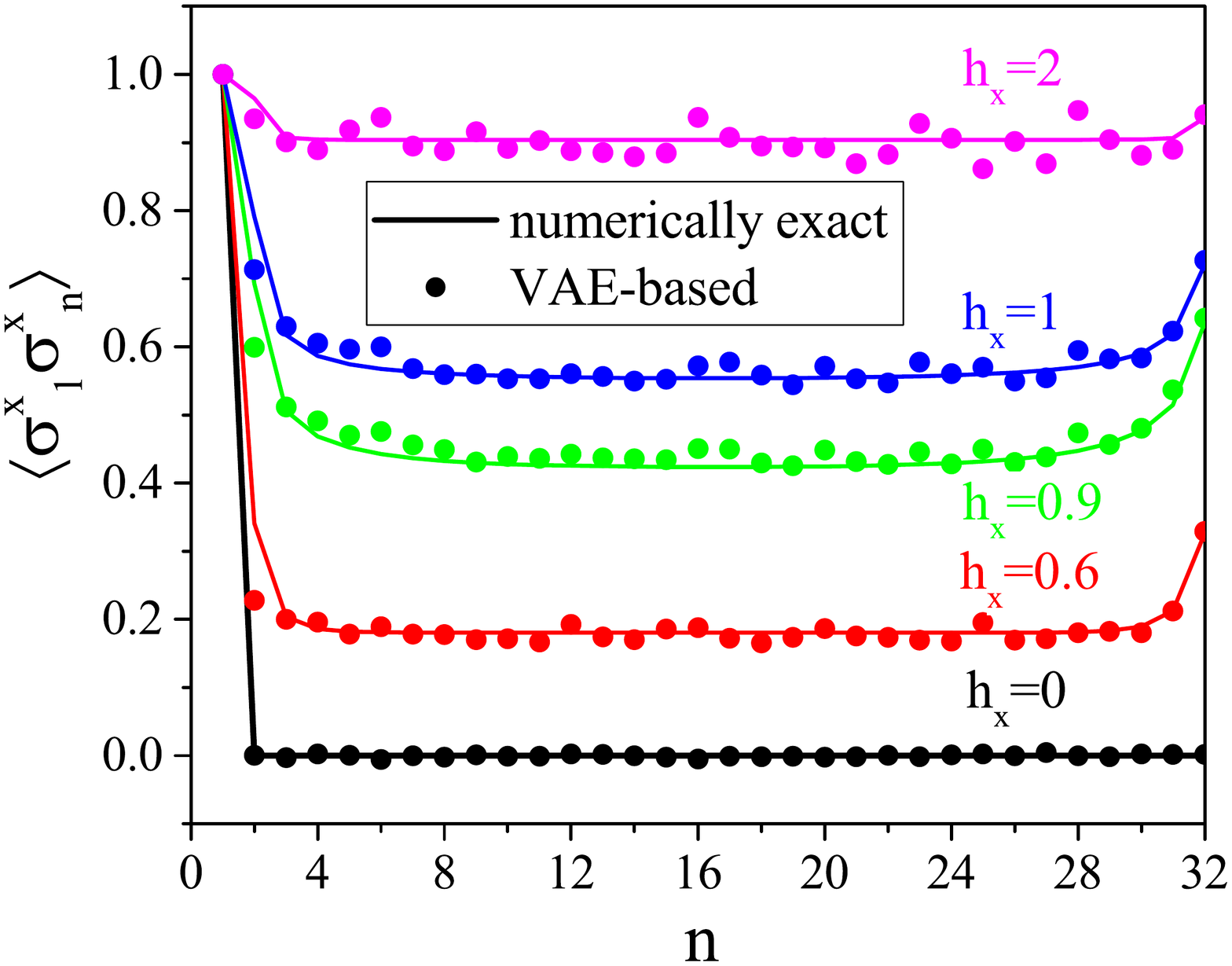}
\caption{Two-point correlation function $\langle \sigma^x_1 \sigma^x_n \rangle$ for different values of external magnetic field $h_x$. }
\label{fig:corr_xx}
\end{figure}

Here, we show that the VAE trained on a set of preliminary measurements is capable to describe the physics of the whole family of TFI models. We validate our results by comparing VAE-based calculations with numerically exact calculations performed by variational MPS algorithm \cite{orus2014practical}. And, to assess the capabilities of the VAE, we consider a spin chain with $32$ spins. We calculate the  MPS representation of the ground state and extract information from it by performing measurements over the state. The external field in the $x$-direction is varied from $0$ to $2$ with a step of $0.1$. The VAE is trained on a data set (TFI measurement outcomes) consisting of $10.5$ million samples in total: $21$ external fields $h_x$ with $500 000$ samples per field.

\begin{figure}[h!]
\centering
\includegraphics[scale=0.4]{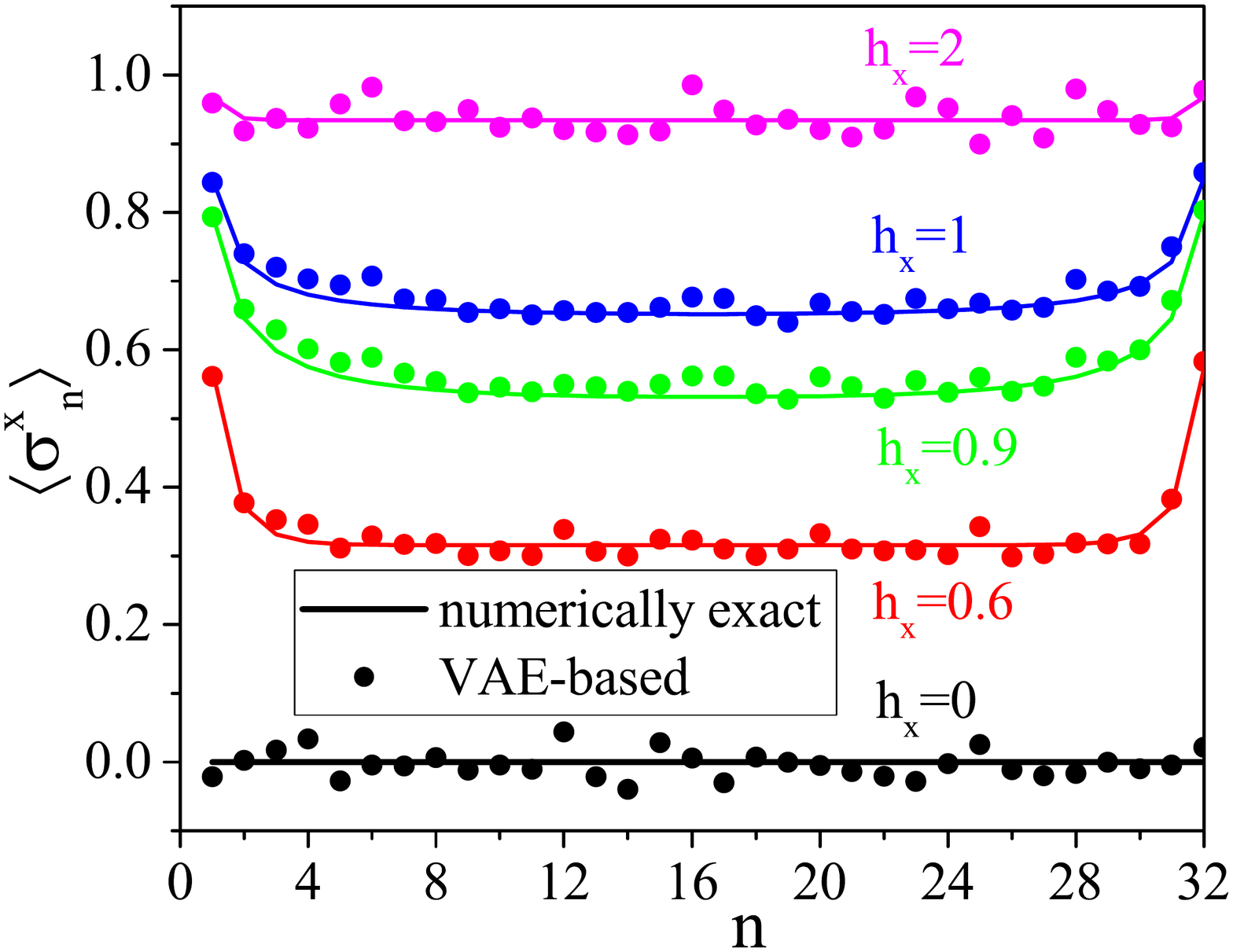}
\caption{Average magnetization per site along $x$ for different values of external magnetic field $h_x$.}
\label{fig:meanx}
\end{figure}

\begin{figure}[h!]
\centering
\includegraphics[scale=0.4]{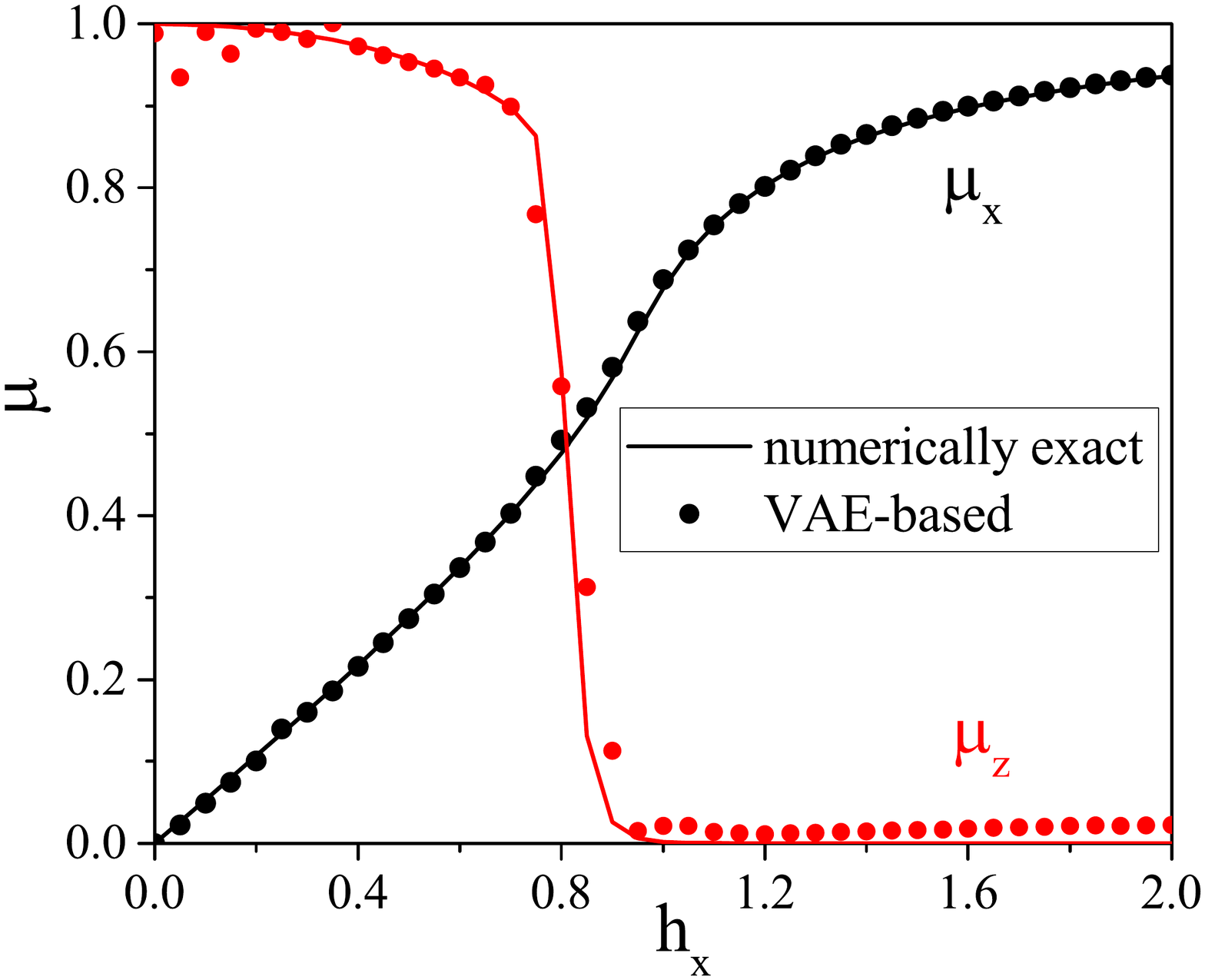}
\caption{Total magnetization along $x$ and $z$ axes for different values of external magnetic field $h_x$. The location of the critical region is slightly shifted towards smaller values of $h_x$ due to the finite size of the chain.}
\label{fig:magn}
\end{figure}

To evaluate the VAE performance, we simply compare directly the numerically exact correlation functions with those reconstructed with our VAE. For $n=1,\ldots,32$, $\langle \sigma^1_z \sigma^n_z \rangle$ and $\langle \sigma^1_x \sigma^n_x \rangle$ are shown in Fig.~\ref{fig:corr_zz} and Fig.~\ref{fig:corr_xx} respectively; and we compare the numerically exact and the VAE-based average magnetizations along $x$, given by $\langle  \sigma^n_x \rangle$ for each position of the spin along the chain, in Fig.~\ref{fig:meanx}. We see that the VAE captures well the physics of the one- and two-point correlation functions. Figure~\ref{fig:magn} shows the total magnetizations, $\mu_x$ and $\mu_z$, in the $x$ and $z$ directions respectively, with $\mu_i = \frac{1}{N}\sum_{j=1}^{N}\langle\sigma_{i}^{j}\rangle$, and we see that the VAE is a tool well-suited for the description of the quantum phase transition and also finite-size effects: while for the infinite TFI chain, i.e. in the thermodynamic limit, the phase transition is observed at $h_x=1$, the finite size of the system yields a shift of the critical point at $h_x \approx 0.9$. Also note that in the $T\rightarrow 0$ limit, the magnetization $M$ defined in Eq. (\ref{magthermal}), coincides exactly with the magnetization $\mu$ defined above.

\begin{figure}[h!]
\centering
\includegraphics[scale=0.4]{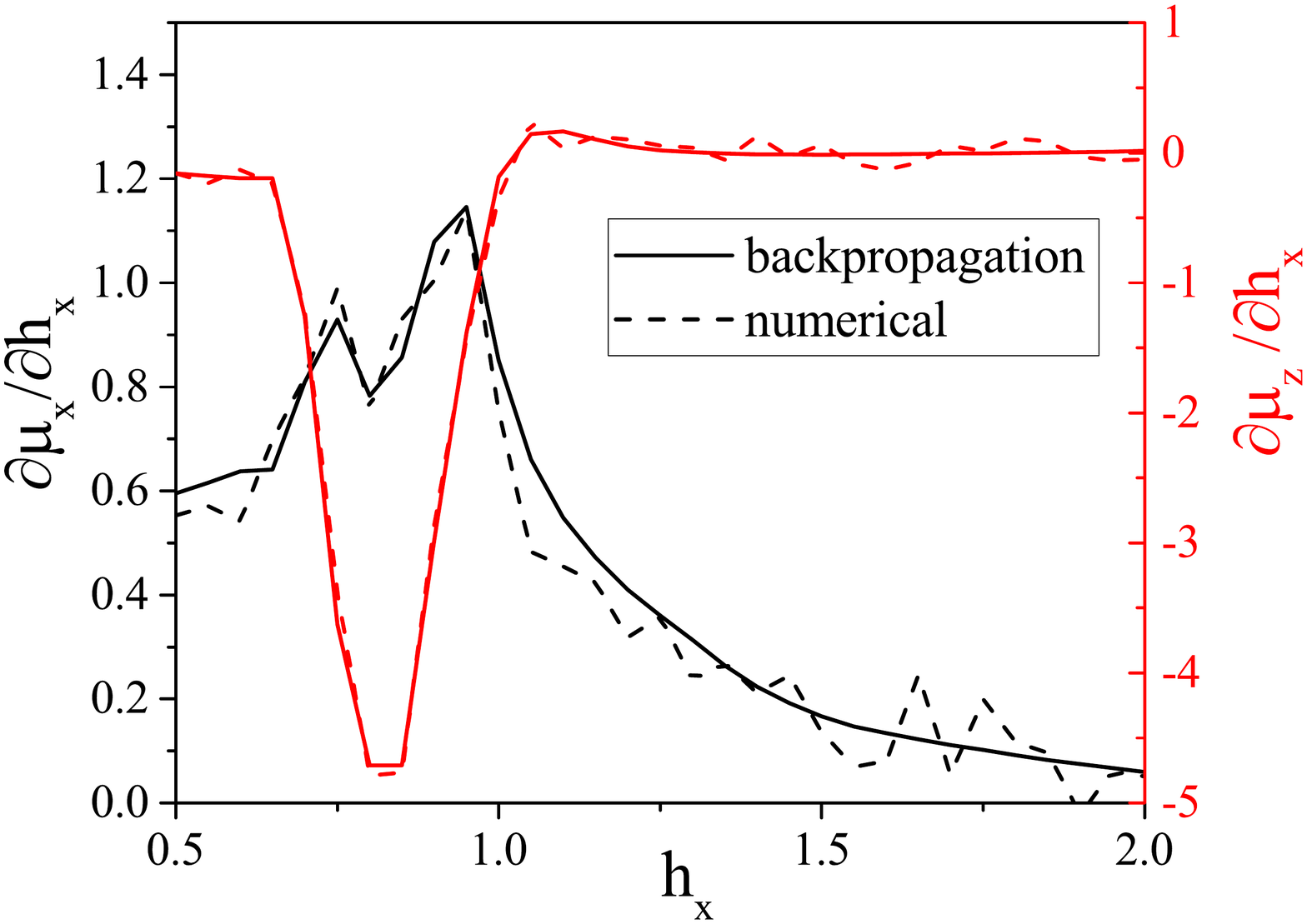}
\caption{Backpropagation-based and numerical-based (central differences) values of $\chi_{xx}$ and $\chi_{zx}$ for different values of external magnetic field $h_x$. Both derivatives slightly fluctuate due to VAE error.}
\label{fig:Mx_grad}
\end{figure}

A back-propagation algorithm combined with the Gumbel-softmax trick may be used to evaluate the derivative of an output over an input. We use this approach to calculate some elements of a magnetic differential susceptibility tensor $\chi_{i j}=\partial \mu_i/\partial h_j$, in particular, $\chi_{x x}$ and $\chi_{z x}$ shown in Fig.~\ref{fig:Mx_grad}. The backpropagation-based magnetic differential susceptibility agrees well with the numerically calculated one (central differences). The main advantage of the backpropagation-based calculation is its numerical efficiency. The VAE may thus be trained with an arbitrary set of external parameters, i.e. not only $h_x$, but also $h_y$ and $h_z$, and yield the full differential susceptibility tensor.  

\begin{figure}[h!]
    \centering
    \includegraphics[scale=0.4]{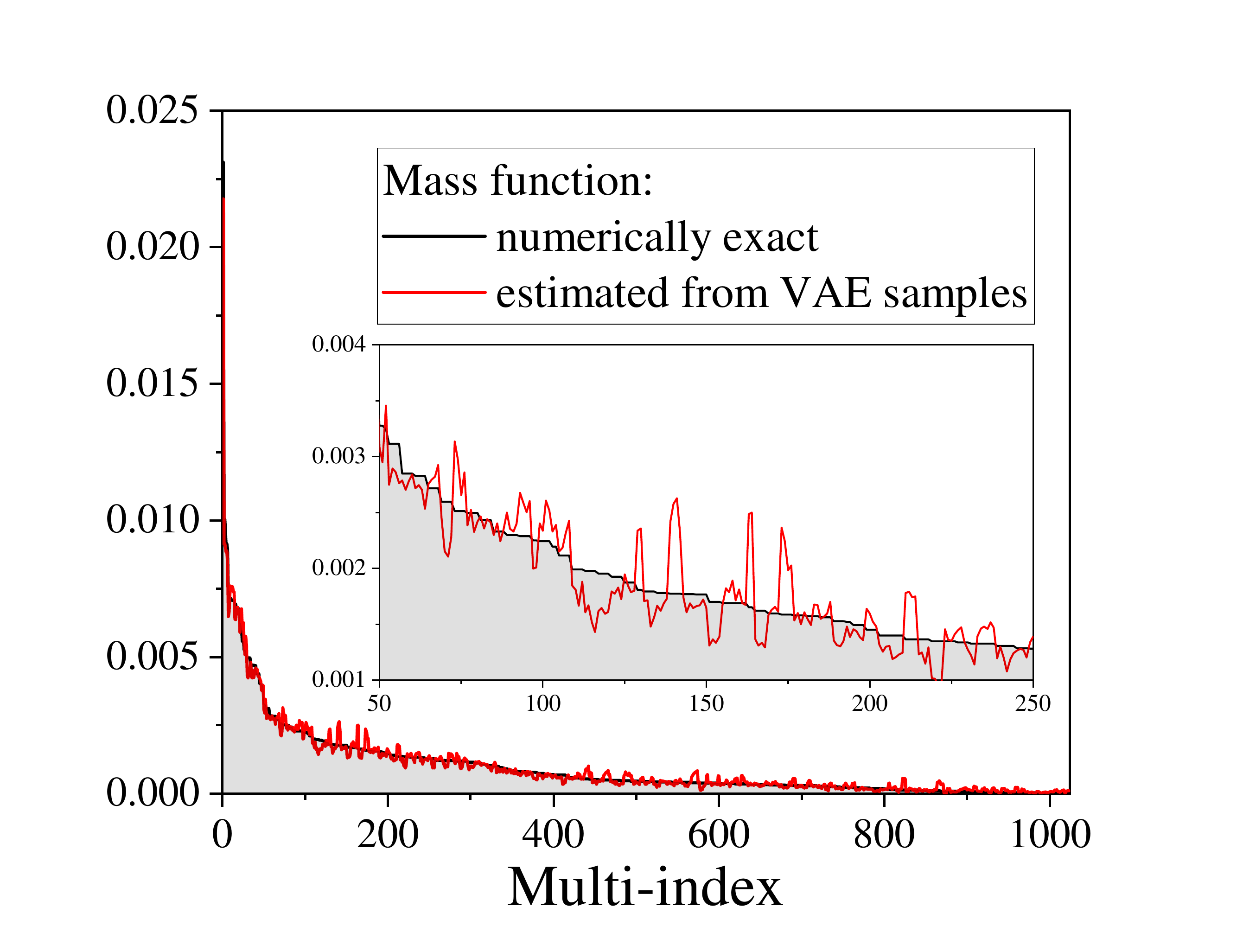}
    \caption{Comparison of two POVM induced mass functions ($P[\alpha] = {\rm Tr}(\rho M^\alpha)$) for a chain of size $5$: numerically exact mass function and reconstructed from VAE samples mass function. A sequence of indices $\alpha$ has been transformed into a single multi-index. Indices have been ordered to put numerically exact probability in descending order. A good agreement between the mass functions is observed.}
    \label{vae_mass_vs_exact_mass}
\end{figure}

\begin{figure}[h!]
    \centering
    \includegraphics[scale=0.4]{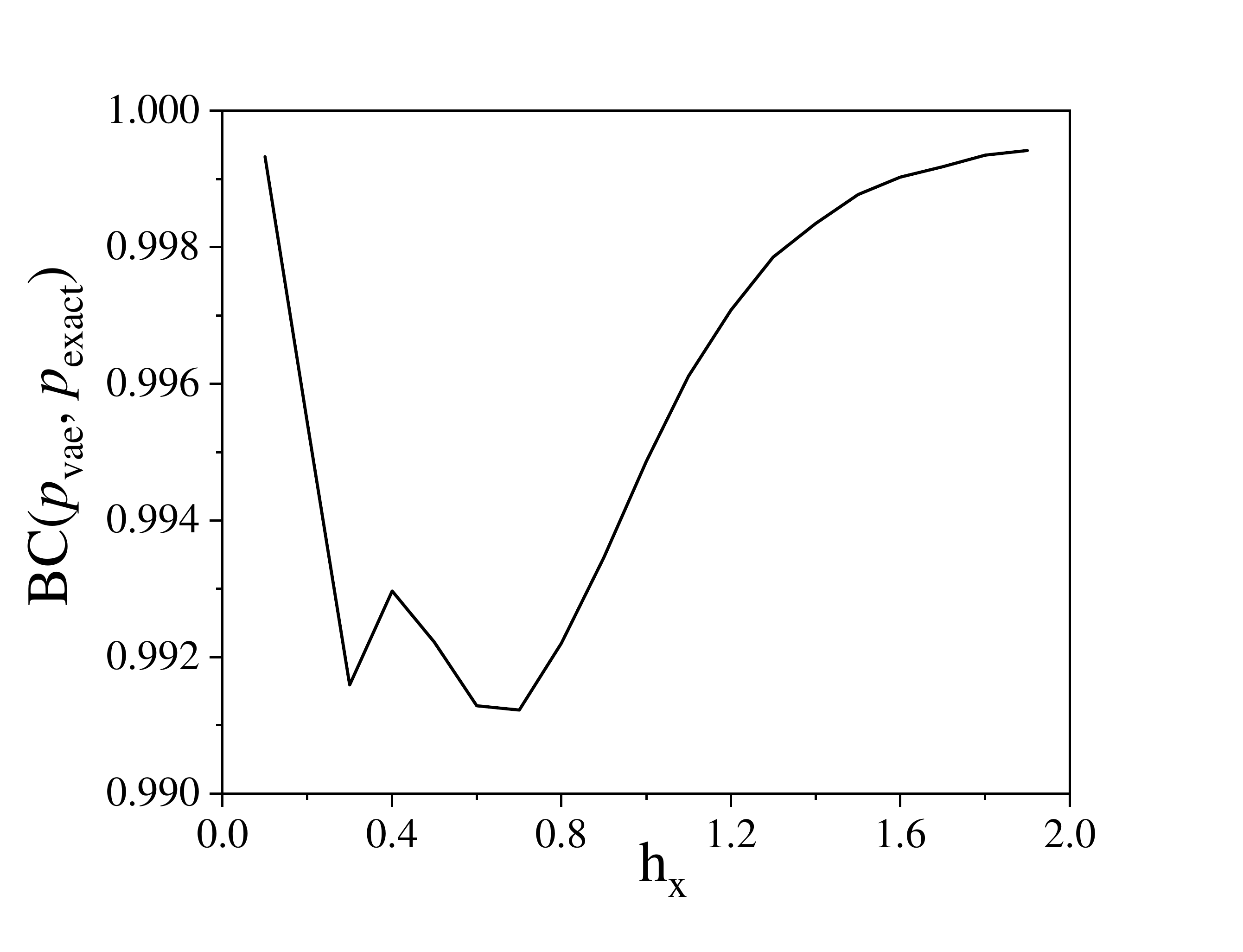}
    \caption{Dependence of the classical fidelity on the external magnetic field. A high predictive accuracy is demonstrated for the whole set of fields.}
    \label{fidelity_vs_field}
\end{figure}

At this stage, we could conclude that the VAE is capable to describe the physics of one- and two-point correlation functions, and hence the TFI physics. However, notwithstanding the ability of the VAE to yield correlation functions that fit well numerically-exact correlation functions, this is not yet a full proof that it represents quantum states well. To address this point, we consider a small spin chain (five spins with TFI Hamiltonian and an external magnetic field $h_x=0.9$) for which we calculate both the exact mass function and that estimated from VAE samples. Figure~\ref{vae_mass_vs_exact_mass} shows that the VAE result again fits the numerically exact mass function with high accuracy. Further, we calculate the Bhattacharyya coefficient \cite{bhattacharyya1943measure}: ${\rm BC}(p_{\rm vae}, p_{\rm exact})=\sum_\alpha p_{\rm exact}[\alpha]\sqrt{\frac{p_{\rm vae}[\alpha]}{p_{\rm exact}[\alpha]}}$  as a function of the external magnetic field $h_x$. Results reported in Fig.~\ref{fidelity_vs_field} show that ${\rm BC}(p_{\rm vae}, p_{\rm exact})>0.99$ over the whole $h_x$ range, which thus proves that the VAE represents a quantum state well, at least for small spin chains.

\begin{figure}[h!]
    \centering
    \includegraphics[scale=0.4]{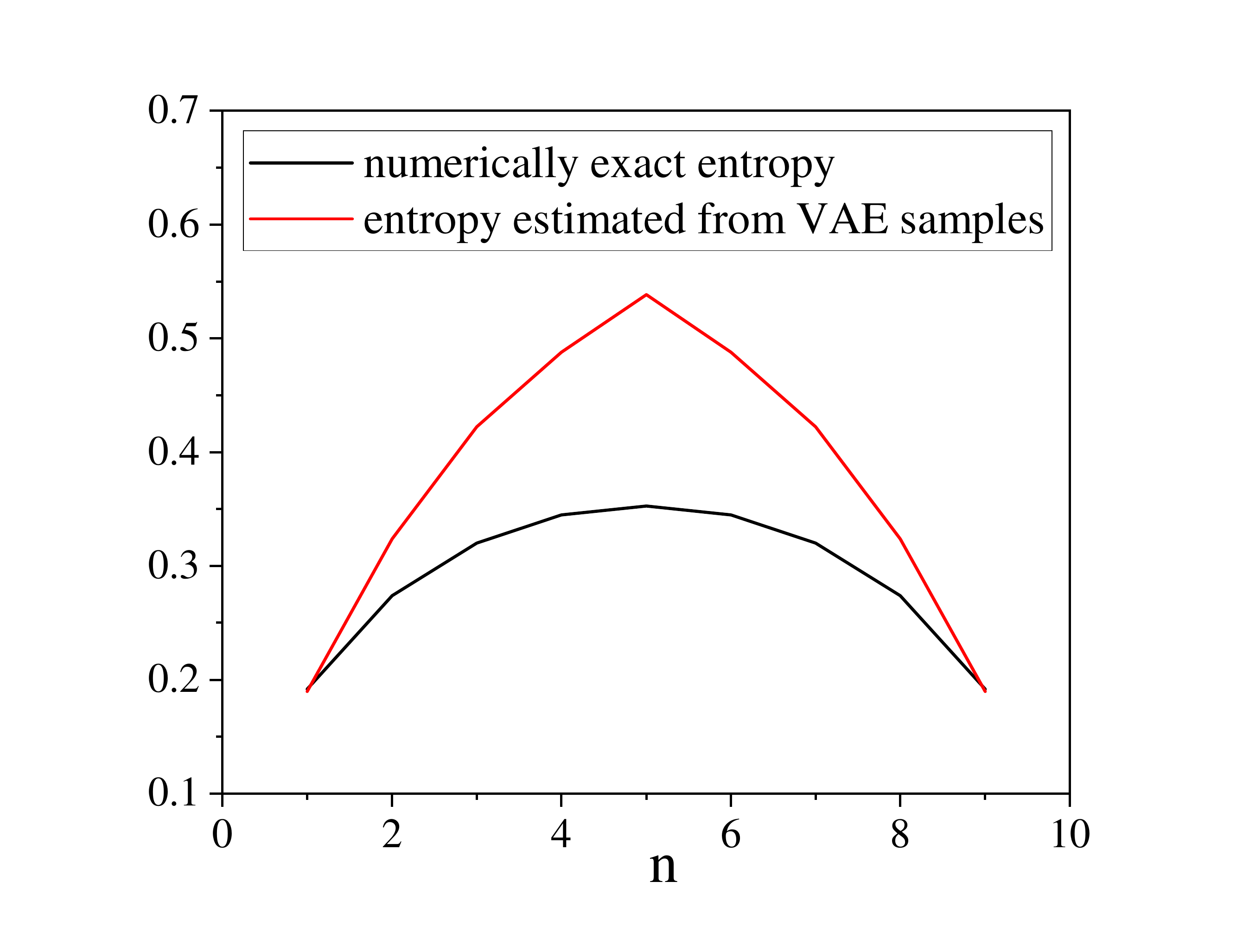}
    \caption{Comparison of the numerically exact R\'enyi entropy and that reconstructed from the VAE samples for different values of $n$.}
    \label{entropy}
\end{figure}

The structure of the entanglement is an another interesting object, we would like to validate. The essence of entanglement between two parts of the chain, which is split into $n$ left spins and $N-n$ right spins, can be described by the R\'eniy entropy of the left part of this chain: $S_\alpha=\frac{1}{1-\alpha}\log{\rm Tr}\rho_{\rm n}^\alpha$, where $\rho_{\rm n}$ is the density matrix of the first $n$ spins in the chain. We estimate the R\'enyi entropy of order $2$: $S_2=-\log({\rm Tr}\rho^2)$, since it can be efficiently calculated from the matrix product representation of the density matrix and from the VAE samples. However, as samples-based estimation of the entangled entropy has a variance which grows exponentially with the number of spins, we consider a small spin chain of size $10$. A direct comparison between the numerically exact and the VAE-based entangled entropies is shown for different values of $n$ in Fig.~\ref{entropy}. For this particular case, the VAE clearly overestimates the entangled entropy. This undesirable effect is in fact, observed for all sizes of spin chains, and even for the spin chain of size $5$ for which we have an excellent agreement between the numerically exact mass function and the VAE-based result. One reason is that $S_2$ is very sensitive to small errors in the mass function, but a deeper understanding of why the VAE overestimates the entangled entropy is the object of future investigation.

\section{Conclusion}
In the present work, we studied the ability of a VAE to reconstruct the physics of quantum many-body systems, using the transverse-field Ising model as a non-trivial example. We used the IC POVM to map the quantum problem onto a probabilistic domain and vice-versa. We trained the VAE on a set of samples from the transformed quantum problem, and our numerical experiments show the following results:

\begin{itemize}
    \item For a large system (32 spins) the VAE's reliability is verified by comparing one- and two-point correlation functions.
    \item For small system (5 spins) the VAE's reliability is verified by direct comparison of mass functions.
    \item The VAE can capture a quantum phase transition.
    \item The response functions (magnetic differential susceptibility tensor) can be obtained using backpropagation through VAE.
    \item Despite the very good agreement between the VAE-based mass function and the true mass function, the VAE shows limited performance with the determination of the entangled entropy. This is point is the object of further development.
\end{itemize}

Our method can be extended to any other thermodynamic system by introduction of the temperature as an external parameter, thereby considering also thermal phase transitions. As one can calculate different thermodynamic quantities by applying backpropagation through VAE, a worthwhile and highly complex system to study would be water under its difference phases to test recent new ideas and models \cite{Artemov2014,Artemov2019}. \\

Our code for our numerical experiments is available on the GitHub repository website \cite{github}.

\acknowledgments
The authors thank Stepan Vintskevich for fruitful discussions. I.A.L. thanks the Russian Foundation for Basic Research for partial support under Project No. 18-37-00282 and Project No. 18-37-20073. S.N.F. acknowledges the Russian Foundation for Basic Research for partial support under Project No. 18-37-20073. A. R. and H. O. acknowledge partial support from the Skoltech NGP Program (Skoltech-MIT joint project).

\appendix
\section{VAE: training and implementation details.}

When training our VAE, we find the arg maximum of the logarithmic likelihood ${\cal L}(\theta)$ w.r.t. its parameters $\theta$:


\begin{eqnarray}
\theta_{\rm MLE} = \underset{\theta}{\rm argmax}{\cal L}(\theta) =  \underset{\theta}{\rm argmax} \log(p[x | \theta, h]),
\label{theta_mle}
\end{eqnarray}

\noindent Equation (\ref{theta_mle}) cannot directly be evaluated, because of hidden variables in the structure of $p[x | \theta, h]$. We can, however, simplify this problem by introducing a distribution over hidden variables $z$. Remember that the probability distribution can be described as $p[x | \theta, h] = \int p[x | z, \theta, h] p[z] dz$, so that the expression for the log likelihood becomes:


\begin{eqnarray}
{\cal L}(\theta)= \log \left ( \int p[x | z, \theta, h] p[z] dz \right ).
\end{eqnarray}

\noindent We can then use a mathematical trick that might seem counterintuitive at first glance, but ultimately becomes quite powerful. We multiply the function inside the integral by $\frac{q[z | x, \Tilde{\theta}, h]}{q[z | x, \Tilde{\theta}, h]}=1$, where $q[z | x, \Tilde{\theta}, h]$ is some arbitrary distribution that can be adjusted with $\Tilde{\theta}$, so that 


\begin{eqnarray}
{\cal L}(\theta) = \log \left ( \int p[x | z, \theta, h] p[z] dz \right ) = \log \left ( \int \frac{q[z | x, \Tilde{\theta}, h]}{q[z | x, \Tilde{\theta}, h]} p[x | z, \theta, h] p[z] dz \right ) \nonumber \\
= \log \left ( \mathbb{E}_{q[z | x, \Tilde{\theta}, h]} p[x | z, \theta, h] \frac{p[z] }{q[z | x, \Tilde{\theta}, h]}\right )
\end{eqnarray}

\noindent where the quantity $\mathbb{E}_{f[x]}$ denotes the expectation value w.r.t some distribution $f[x]$. We can then use Jensen's inequality to show that 

\begin{eqnarray}
\log \left ( \mathbb{E}_{q[z | x, \Tilde{\theta}, h]} p[x | z, \theta, h] \frac{p[z] }{q[z | x, \Tilde{\theta}, h]}\right ) \geq \mathbb{E}_{q[z | x, \Tilde{\theta}, h]} \log \left ( p[x | z, \theta, h] \frac{p[z] }{q[z | x, \Tilde{\theta}, h]}\right ).
\end{eqnarray}

\noindent where the rhs of this inequality is the lower bound of the log likelihood since it will always be greater than or equal to the lower bound, and equality can always be achieved by a proper choice of $q$ if it is in a complex enough family. 

Maximizing the lower bound is equivalent to maximizing the log likelihood. We can decompose this lower bound term into two terms:

\begin{eqnarray}
{\cal L}(\theta) \geq \mathrm{ELBO}(\theta, \Tilde{\theta}) = \mathbb{E}_{q[z | x, \Tilde{\theta}, h]} \log \left ( p[x | z, \theta, h] \right ) - \int q[z | x, \Tilde{\theta}, h] \log \frac{q[z | x, \Tilde{\theta}, h]}{p[z]}dz
\end{eqnarray}

\noindent Note that the second term is equivalent to the Kullback-Leibler divergence $KL(q[z | x, \Tilde{\theta}, h] \, || \, p[z])$. In our case, we picked the particular distribution forms, which reflect the structure of our problem:

\begin{eqnarray}
 p[x | z, \theta, h] = \prod_{i=1}^N \prod_{j=1}^4 \pi_{ij}(z, \theta, h)^{x_{ij}}, \nonumber \\
 q[z|x, \Tilde{\theta}, h] = {\cal N} (\mu_i(x, \Tilde{\theta}, h), \mathrm{Diag}(\sigma_i^2 (x, \Tilde{\theta}, h))), \nonumber \\
 P[z] = {\cal N}(0, I)
\end{eqnarray}

\noindent where $\mu_i$ and $\sigma_i$ are given by the encoder neural network, and $\pi_{ij}$ is given by the decoder neural network, with $\sum_{j=1}^4 \pi_{ij} = 1$ and $\pi_{ij} \geq 0$, which can be achieved by applying the softmax funtion to the output of the neural network. Now we can use the reparametrization trick to change the variable in the integral $z = \sigma_j (x, \Tilde{\theta}, h) \varepsilon + \mu_j (x, \Tilde{\theta}, h)$ where $\varepsilon_j \sim {\cal N} (0,I)$ to simplify this expression to:

\begin{eqnarray}
\mathrm{ELBO}(\theta, \Tilde{\theta}) =  \sum_{i=1}^N \sum_{j=1}^4 x_{ij} \left \langle \log \left ( \pi_{ij}(\sigma_i(x, \Tilde{\theta}, h) \varepsilon + \mu_i(x, \Tilde{\theta}, h), \theta, h) \right ) \right \rangle_{\varepsilon_j \sim {\cal N} (0,I)} \nonumber \\
- \sum_{i=1}^N \left ( \log \sigma_i(x, \Tilde{\theta}, h) - \frac{\sigma_i^2(x, \Tilde{\theta}, h) + \mu_i^2(x, \Tilde{\theta}, h) -1}{2} \right ).
\label{eq:ELBO}
\end{eqnarray}

\begin{figure}
\centering
\resizebox{0.5\textwidth}{!}{\includegraphics{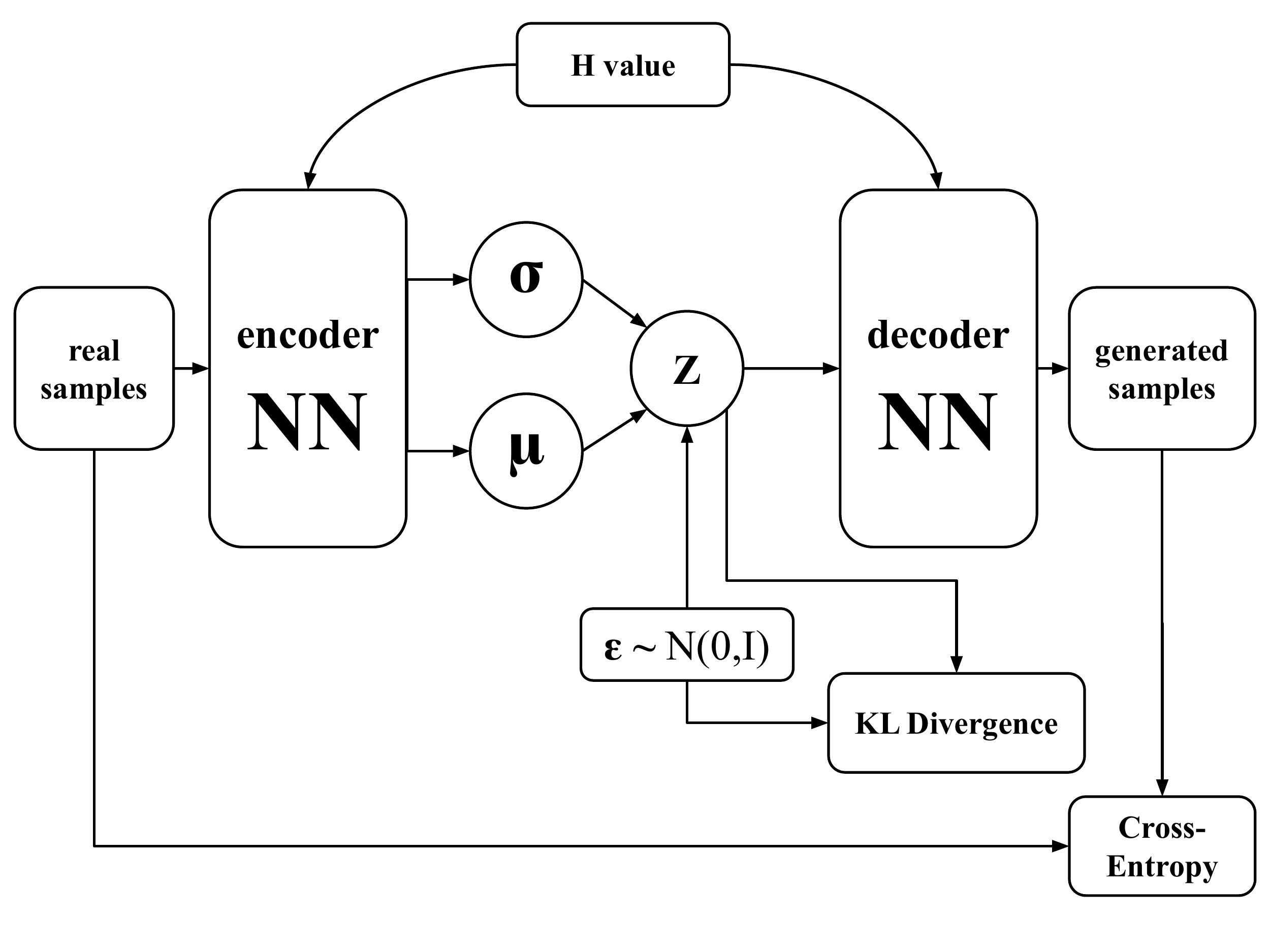}}
\caption{Architecture of the Variational Autoencoder}
\label{fig:vae_architecture}
\end{figure}

The first term is the cross-entropy, which pushes the probability distribution to be as close as possible to the data. The second term is the regularizer, which forces the latent variable $z$ not to diverge too much from the normal distribution ${\cal N} (0,I)$ so that the VAE can be used to generate new data once it is trained. Note that both $x_{ij}$ and $\sigma_i$ must be positive. Instead of adding a constraint to the VAE, which would be difficult to do, we train the VAE for the variables $\Pi = \log \pi$ and $\xi = 2\log \sigma$. Equation (\ref{eq:ELBO}) then becomes:

\begin{eqnarray}
\mathrm{ELBO}(\theta, \Tilde{\theta}) = \sum_{i=1}^N \sum_{j=1}^4 x_{ij} \left \langle \Pi_{ij}(e^{\xi_i(x, \Tilde{\theta}, h) /2} \varepsilon + \mu_i(x, \Tilde{\theta}, h), \theta, h)  \right \rangle_{\varepsilon_j \sim {\cal N} (0,I)} \nonumber \\
- \frac{1}{2} \sum_{i=1}^N \left ( \xi_i(x, \Tilde{\theta}, h) -e^{\xi_i(x, \Tilde{\theta}, h)} - \mu_i^2(x, \Tilde{\theta}, h) + 1 \right ).
\end{eqnarray}

Now, $\mathrm{ELBO}(\theta, \Tilde{\theta})$ can be effectively optimized using gradient descent methods, averaging over $\varepsilon$ can be done by sampling. Generalizing to a data set of size $M$: $\{x^k\}_{k=1}^M$ can be easily done and is shown by:

\begin{eqnarray}
\mathrm{ELBO}(\theta, \Tilde{\theta}) = \sum_{k=1}^M \sum_{i=1}^N \sum_{j=1}^4 x_{ij}^k \left \langle \Pi_{ij}(e^{\xi_i(x^k, \Tilde{\theta}, h) /2} \varepsilon + \mu_i(x^k, \Tilde{\theta}, h), \theta, h)  \right \rangle_{\varepsilon_j \sim {\cal N} (0,I)} \nonumber \\
- \frac{1}{2} \sum_{k=1}^M  \sum_{i=1}^N \left ( \xi_i(x^k, \Tilde{\theta}, h) -e^{\xi_i(x^k, \Tilde{\theta}, h)} - \mu_i^2(x^k, \Tilde{\theta}, h) + 1 \right ).
\end{eqnarray}

\noindent A visual representation of the VAE architecture is shown in Fig. \ref{fig:vae_architecture}.

To solve the optimization problem we use Adam optimizer \cite{kingma2014adam} with standard parameters (${\rm lr}=0.001, \beta_1=0.9, \beta_2=0.999$). For the encoder and decoder we use fully-connected neural networks with $2$ hidden layers and $256$ neurons on each. We train the VAE using batches of size $100 000$ samples and for $750$ epochs.

\section{Sampling from POVM induced mass function}

The mass function induced by POVM $P[\alpha_1, \alpha_2,\dots,\alpha_N]$ has a form of matrix product state. Thus, one can easily calculate any marginal mass function because a summation over any $\alpha$ can be done locally. Any conditional mass functions can be also calculated by using marginal mass functions. Thus, one can calculate chain decomposition of the whole mass function:

\begin{equation}
P[\alpha_1, \alpha_2,\dots,\alpha_N] = 
P[\alpha_N]P[\alpha_{N-1}|\alpha_N]P[\alpha_{N-2}|\alpha_{N-1},\alpha_N]\dots P[\alpha_1|\alpha_2,\dots,\alpha_N]
\end{equation}

\noindent With this decomposition one can produce a sample $\Tilde{\alpha}_N$ from $P[\alpha_N]$ first, then a sample $\Tilde{\alpha}_{N-1}$ from $P[\alpha_{N-1}|\Tilde{\alpha}_N]$ and continue up to the end of the chain. The obtained set $\{\Tilde{\alpha}_1, \Tilde{\alpha}_2, \dots, \Tilde{\alpha}_N\}$ is a valid sample from the mass function.

\section{Abbreviations}

The following abbreviations are used in this manuscript:\\

\begin{tabular}{@{}ll}
VAE & Variational Autoencoder\\
MPS & Matrix product state\\
TFI & Transverse-field Ising\\
IC & Informationally incomplete\\
POVM & Positive-operator valued measure\\
ELBO & Evidence lower bound\\
NN & Neural network\\
KL &  Kullback–Leibler\\
DMRG & Density matrix renormalization group
\end{tabular}

~\\

\bibliography{bibliography}

\end{document}